\documentclass[a4paper, 11pt]{article}
\usepackage{jheppub,multirow,morefloats}

\title{\boldmath Heavy Higgs signal-background interference in $gg\to VV$ in the Standard Model plus real singlet}
\author{Nikolas Kauer and Claire O'Brien}
\affiliation{Department of Physics, Royal Holloway, University of London, Egham Hill, Egham TW20 0EX, U.K.}
\emailAdd{n.kauer@rhul.ac.uk}
\emailAdd{claire.obrien.2012@live.rhul.ac.uk}

\abstract{
For the Standard Model extended with a real scalar singlet field, the modification 
of the heavy Higgs signal due to interference with the continuum background and the
off-shell light Higgs contribution is studied for $gg\to ZZ, WW \to 4$ lepton 
processes at the Large Hadron Collider. 
Interference effects can range from $\calO(10\%)$ to 
$\calO(1)$ effects for integrated cross sections.  
Despite a strong cancellation between the heavy Higgs-continuum 
and the heavy Higgs-light Higgs interference, 
the full interference is clearly non-negligible and modifies
the heavy Higgs line shape.
A $\left|M_{VV} - M_{h2}\right| < \Gamma_{h2}$ cut mitigates 
interference effects to $\calO(10\%)$ or less.
A public program that allows to simulate 
the full interference is presented.
}


\newcommand{\sla}[1]{\ifmmode%
  \setbox0=\hbox{$#1$}%
  \setbox1=\hbox to\wd0{\hss$/$\hss}\else%
  \setbox0=\hbox{#1}%
  \setbox1=\hbox to\wd0{\hss/\hss}\fi%
  #1\hskip-\wd0\box1 }
  \newcommand{\ea}[1]{\begin{align}#1\end{align}}
\newcommand{\calM}{{\cal M}}
\newcommand{\calO}{{\cal O}}

\begin{document}


\maketitle


\section{Introduction}
In 2012, the ATLAS and CMS experiments at the CERN Large Hadron 
Collider (LHC) announced the discovery of a new scalar resonance with a mass of 
approximately 125 GeV \cite{HiggsExperiment}.
The discovered particle is so far consistent with the Higgs 
boson predicted by the Standard Model (SM) Higgs mechanism 
\cite{HiggsTheory}, but many extensions to the SM preserve the minimal 
assumptions of an $SU(2)$ doublet which acquires a vacuum expectation value 
thus inducing a physical Higgs boson that couples to fermions and vector bosons in 
proportion to their mass, while also allowing for an expanded Higgs sector with 
additional, heavier (or lighter) Higgs-like scalar particles. The search for 
high-mass Higgs-like particles in the $gg\to H\to ZZ$ and $gg\to H\to WW$ channels 
at the LHC is ongoing \cite{Aad:2012oxa,Aad:2012me,ATLAS:2012ac,Chatrchyan:2013yoa,Chatrchyan:2012dg,Chatrchyan:2012sn,CMS:2013cda,Diglio:2014vpa,Khachatryan:2015cwa,Pelliccioni:2015hva}.

With inclusive NNLO signal uncertainties of $\mathcal{O}(10\%)$ in 
gluon-fusion Higgs production at the LHC, which can be further reduced by 
experimental selection cuts, it is important to study signal-background interference 
in the $H\to VV$ decay modes ($V=Z,W$), because it can be of similar size or larger 
for Higgs invariant masses above the weak-boson pair threshold.  For Higgs 
invariant masses much larger than $2M_V$, the occurring sizeable Higgs-continuum 
interference is linked to the preservation of unitarity.
In the SM, interference between the Higgs signal and continuum background in 
$gg\ (\to H)\to VV$ and including fully leptonic decays has been studied in refs.\ \cite{Glover:1988fe,Glover:1988rg,Binoth:2006mf,Campbell:2011cu,Kauer:2012ma,Passarino:2012ri,Kauer:2012hd,Bonvini:2013jha,Kauer:2013qba,Campbell:2013una,Moult:2014pja,Ellis:2014yca,Campanario:2012bh,Campbell:2014gua,Li:2015jva}.\footnote{We note that the interfering $gg\to VV$ continuum background at LO is 
formally part of the NNLO corrections to $pp\to VV$ \cite{Cascioli:2014yka,Gehrmann:2014fva}.
SM Higgs-continuum interference in the $H\to VV$ decay modes at a $e^+e^-$
collider has been studied in ref.\ \cite{Liebler:2015aka}.
Predictions for $gg\to\ell\ell\nu\nu+0,1$ jets have been presented in ref.\ \cite{Cascioli:2013gfa}.}  
Higgs-continuum interference results for a heavy SM Higgs boson with a $\Gamma_H/M_H$ ratio of $\mathcal{O}(10\%)$ or more have been presented in refs.\ \cite{Campbell:2011cu,Kauer:2012ma,Passarino:2012ri,Campanario:2012bh,Bonvini:2013jha,Kauer:2013qba,Moult:2014pja}.  We note that all 
Higgs-continuum interference calculations are at leading order (LO), except for 
refs.\ \cite{Bonvini:2013jha,Moult:2014pja,Li:2015jva}, where approximate higher-order corrections have been  
calculated.

Since a Higgs boson with $M_H\approx 125$ GeV has been discovered, a theoretically 
consistent search for an additional Higgs boson has to be based on a model 
that is beyond the SM.  The simplest extension of the Higgs sector of the 
SM introduces an additional real scalar singlet field which is neutral under the 
SM gauge groups. The remaining viable parameter space of this 1-Higgs-Singlet
extension of the SM (abbreviated by 1HSM) after LHC Run 1 has been 
studied in refs.\ \cite{Pruna:2013bma,Robens:2015gla}.%
\footnote{See also refs.\ \cite{Falkowski:2015iwa,Lopez-Val:2014jva,Martin-Lozano:2015dja}.}
Here, we focus on the case where the 
additional Higgs boson is heavier than the discovered Higgs boson.  In this case,
the heavy Higgs signal is affected not only by sizeable interference with the 
continuum background, but also by a non-negligible interference with the off-shell 
tail of the light Higgs boson \cite{Kauer:2012hd}.  
A calculation including full interference effects in a Higgs portal model has 
been carried out in ref.\ \cite{Englert:2014ffa}. But, the occurring interference 
effects (which are discernible in the distributions shown in figure 8 of ref.\ \cite{Englert:2014ffa}) have not been analysed quantitatively there.%
\footnote{We note that we presented preliminary results which demonstrate the 
importance of heavy-light and heavy-continuum interference in September 2014 at 
the HP2 Workshop, Florence.} 
A dedicated study of heavy Higgs-light Higgs 
interference in the 1HSM with an additional $Z_2$ symmetry was 
presented in ref.\ \cite{Maina:2015ela}.\footnote{%
For Higgs production in vector boson fusion, heavy-light interference in a 
two-Higgs model was studied in ref.\ \cite{Liebler:2015aka} for an $e^+e^-$ collider
and in more detail including heavy-continuum interference in ref.\ \cite{Ballestrero:2015jca} for the LHC.}

In this paper, we extend the analysis of ref.\ \cite{Maina:2015ela}
by taking into account the full signal-background interference which
includes the heavy Higgs-continuum interference.%
\footnote{A similar study which numerically agrees with ours has subsequently appeared on the arXiv \cite{Englert:2015zra}.}
Furthermore, in 
addition to $gg\to h_2\to ZZ\to 4$ leptons, where $h_2$ is the 
heavy Higgs boson, 
we also calculate results for $gg\to h_2\to WW\to 4$ leptons.
Our calculations 
are carried out with a new version of the parton-level integrator and event generator \textsf{gg2VV}, which 
we have made publicly available \cite{gg2VVurl}.
In section \ref{sec:model} we discuss the 1HSM and specify the used benchmark points.
Calculational details are discussed in section \ref{sec:calc}. Integrated cross 
sections and differential distributions in $M_{VV}$ for the heavy Higgs 
signal and its interference with the continuum background and off-shell 
light Higgs contribution are presented in section 
\ref{sec:results} for 
$g g \to h_2 \to Z Z \to \ell \bar{\ell}\ell' \bar{\ell}'$
and $g g \to h_2 \to W^- W^+ \to \ell \bar{\nu}\bar{\ell}' \nu'$.
Conclusions are given in section \ref{sec:sum}.


\section{Model\label{sec:model}}

As minimal theoretically consistent model with two physical Higgs bosons, we 
consider the SM with an added real singlet field which is neutral under all 
SM gauge groups.
The 1-Higgs-Singlet Extension of the SM has been extensively explored in the literature \cite{Binoth:1996au,Schabinger:2005ei,Patt:2006fw,Bowen:2007ia,Barger:2007im,Barger:2008jx,Bhattacharyya:2007pb,Dawson:2009yx,Bock:2010nz,Fox:2011qc,Englert:2011yb,Englert:2011us,Batell:2011pz,Englert:2011aa,Gupta:2011gd,Dolan:2012ac,Batell:2012mj,No:2013wsa,Profumo:2014opa,Logan:2014ppa,Chen:2014ask}. Higgs singlet models with an additional $Z_2$ symmetry have generated some interest recently because of the possibility of the additional Higgs boson being a dark matter candidate, but here we consider the most general extension. In the following, we give a brief summary of the model.  A more detailed description can be found in refs.\ \cite{Chen:2014ask,Heinemeyer:2013tqa}.

The SM Higgs sector is extended by the addition of a new real scalar field, which is a singlet under all the gauge groups of the SM and which also gets a vacuum expectation value (VEV) under electroweak symmetry breaking. The most general gauge-invariant potential can be written as \cite{Schabinger:2005ei,Bowen:2007ia}
\begin{equation}
V = \lambda \left( \Phi^\dagger\Phi - \frac{v^2}{2}\right)^2 + \frac{1}{2} M^2 s^2
+ \lambda_1 s^4 +\lambda_2 s^2 \left( \Phi^\dagger\Phi - \frac{v^2}{2}\right) 
+\mu_1 s^3 + \mu_2 s \left( \Phi^\dagger\Phi - \frac{v^2}{2}\right), 
\label{genpot}
\end{equation}
where $s$ is the real singlet scalar which is allowed to mix with the SM $SU(2)$ Higgs doublet, which in the unitary gauge can be written as
\begin{equation}
\Phi = \begin{pmatrix}
0 \\
(\phi + v)/\sqrt{2}
\end{pmatrix}
\label{capphi}
\end{equation}
with VEV $v\simeq 246$ GeV.
Here it has already been exploited that (without the $Z_2$ symmetry) shifting the singlet field simply corresponds to a redefinition of the parameter coefficients and 
due to this freedom one can take the VEV of the singlet field to zero, which implies 
$M^2>0$. To avoid vacuum instability the quartic couplings must satisfy 
\begin{equation}
\lambda>0,\quad \lambda_1 > 0,\quad \lambda_2 > -2 \sqrt{\lambda \lambda_1}\,.
\end{equation}
The trilinear couplings $\mu_1$ and $\mu_2$ can have positive or negative sign.
Substituting eq.\ (\ref{capphi}) into eq.\ (\ref{genpot}), one obtains the potential
\begin{equation}
V = \frac{\lambda}{4}\phi^4 + \lambda v^2 \phi^2 + \lambda v \phi^3 + \frac{1}{2}M^2s^2 + \lambda_1 s^4 + \frac{\lambda_2}{2}\phi^2 s^2 + \lambda_2 v \phi s^2 + \mu_1 s^3 + \frac{\mu_2}{2}\phi^2 s + \mu_2 v \phi s\,.
\end{equation}
The mass eigenstates can be parametrised in terms of a mixing angle $\theta$ as
\begin{align}
h_1 &= \phi \cos \theta  - s \sin \theta \,, \\
h_2 &= \phi \sin \theta + s \cos \theta \,,
\end{align}
where $h_1$ is assumed to be the lighter Higgs boson with a mass of approximately 125 GeV, and
\begin{equation}
\tan 2\theta = \frac{- \mu_2 v}{\lambda v^2 - \frac{1}{2} M^2}
\end{equation}
with 
\begin{equation}
-\frac{\pi}{4} < \theta < \frac{\pi}{4}
\end{equation}
under the condition $M^2>2\lambda v^2$.
The model has six independent parameters, which we choose to be $M_{h1}, M_{h2}, \theta, \mu_1, \lambda_1$ and $\lambda_2$.  The dependent model parameters are:
\begin{align}
\lambda &=\frac{\cos\left(2\theta\right)\left(M_{h1}^2 - M_{h2}^2\right) + M_{h1}^2 + M_{h2}^2 }{4v^2}\,, \\
M^2 &= \frac{M_{h2}^2 - M_{h1}^2 + \sec\left(2\theta\right) \left(M_{h1}^2 + M_{h2}^2\right)}{2\sec\left(2\theta\right)}\,, \\
\mu_2 &= -\tan\left(2\theta\right)\ \frac{\lambda v^2 - \frac{1}{2}M^2}{v}\,. 
\end{align}

We set $M_{h1}$ to 125 GeV in accordance with the mass of the observed resonance
and study three values for the mass of the heavy Higgs resonance: $M_{h2}=300$ GeV, $M_{h2}=600$ GeV and $M_{h2}=900$ GeV.
We choose the mixing angle $\theta$ so as not to alter the predicted 
light Higgs cross section too much.
To illustrate how interference effects change with the mixing angle, 
we study the two values $\theta=\pi/15$ and $\theta=\pi/8$, which is consistent with current limits on the Higgs signal strength and does not appear to be in conflict with limits given in ref.\ \cite{Robens:2015gla}, but strictly speaking these apply to the model with the additional $Z_2$ symmetry and are not directly applicable here.
Furthermore, we consider model benchmark points with vanishing coupling parameters $\mu_1, \lambda_1$ and $\lambda_2$. ($\lambda_1>0$ is treated as approximately zero.)  We emphasise that this does not imply that the $h_2\to h_1 h_1$ decay width 
is zero.  For instance, for the mixing angle $\theta=\pi/8$ and $M_{h2}=$ 300 (600) [900] GeV the branching ratio $\Gamma(h_2\to h_1 h_1)/\Gamma_{h2}$ is 28\% (20\%) [19\%].  The $h_2\to h_1 h_1$ decay mode is
therefore not suppressed in our study.  Furthermore, the implementation in \textsf{gg2VV} is not restricted to benchmark points with vanishing $\mu_1, \lambda_1$ and $\lambda_2$.  Nonzero values of $\mu_1, \lambda_1$ and $\lambda_2$ affect the 
calculation of the signal-background interference only via a change of the heavy 
Higgs width.  In combination with \textsf{FeynRules}, our implementation in 
\textsf{gg2VV} therefore allows to calculate full signal-background interference
effects for arbitrary benchmark points of the general 1-Higgs-Singlet Extension of 
the SM.  See section \ref{sec:calc} for further details.
Ref.\ \cite{Chen:2014ask} gives bounds on the $\lambda_1$ and $\mu_1$ parameters for $M_{h2} \lesssim 500$ GeV and a similar $\theta$, which are in agreement with our choice of zero for these parameters.  
Our choice for the coupling parameters is also in agreement with upper limits 
on the combination of these parameters from experimental searches \cite{CMS:2014ipa,CMS:2014eda}.


\section{Calculational details \label{sec:calc}}

\begin{figure}[tb]
\centering
\includegraphics[width=0.48\textwidth, clip=true]{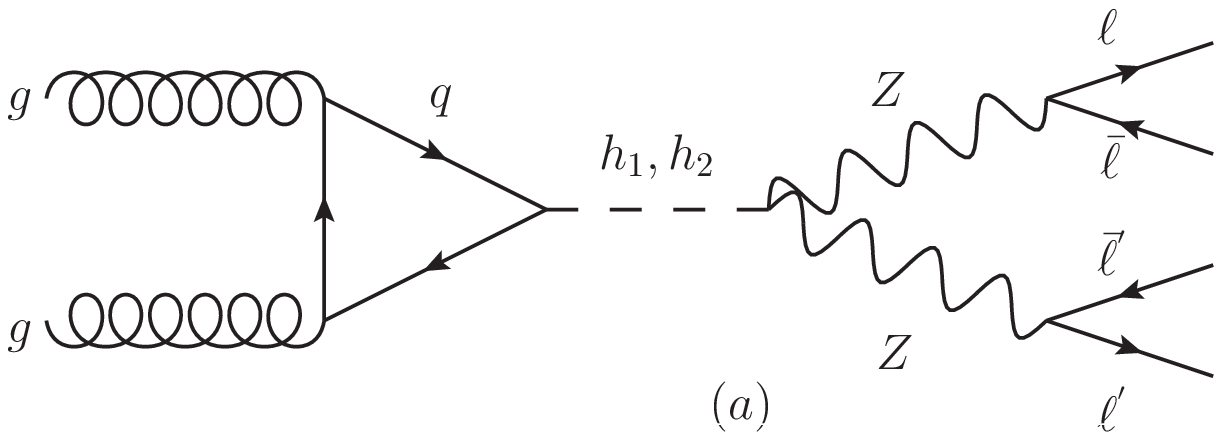}\hfill
\includegraphics[width=0.48\textwidth, clip=true]{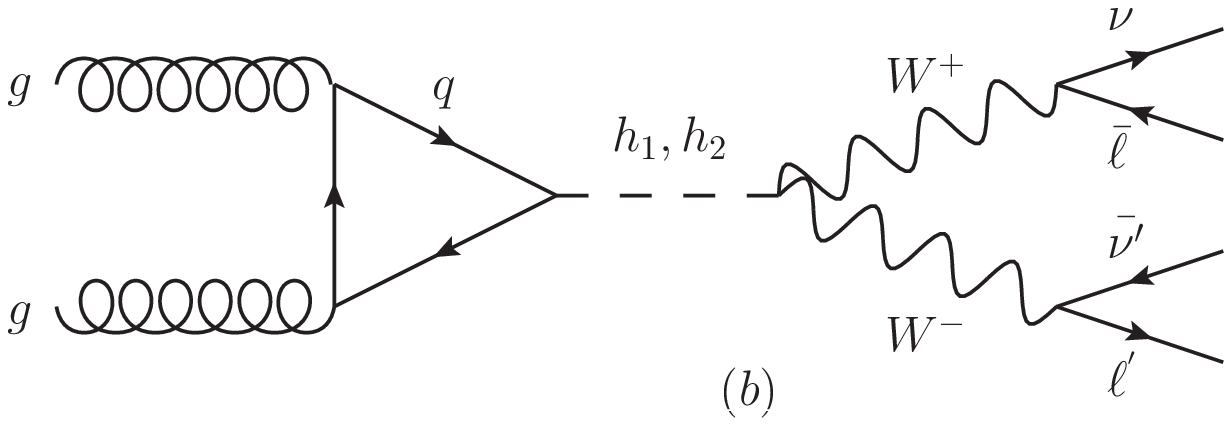}\hfill
\includegraphics[width=0.48\textwidth, clip=true]{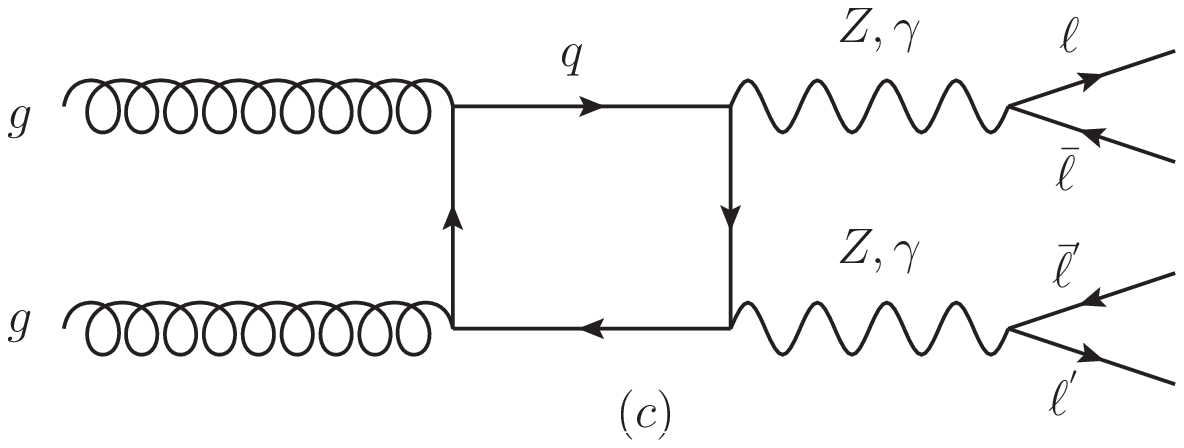}\hfill
\includegraphics[width=0.48\textwidth, clip=true]{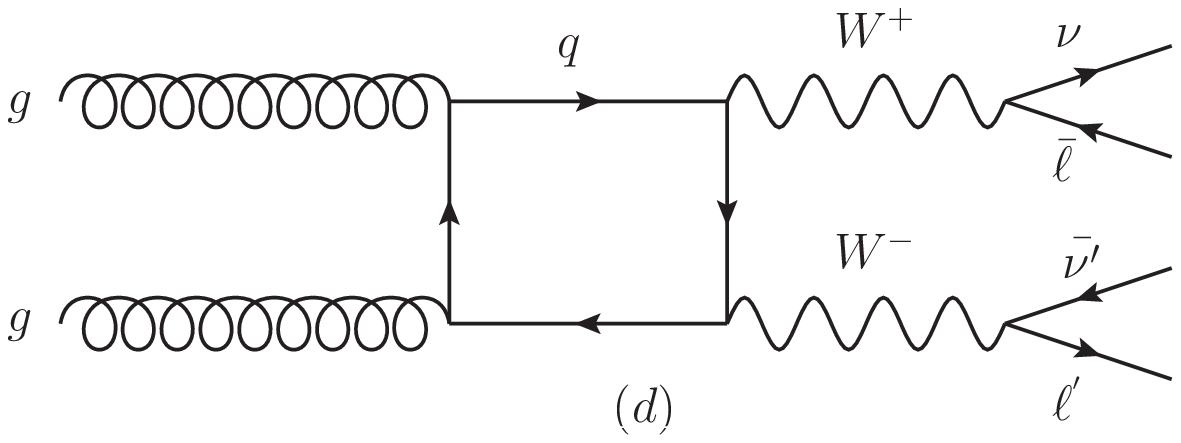}\hfill
\caption{\label{fig:diagrams}Representative Feynman graphs for $g g\ (\to \{h_1,h_2\}) \to ZZ,WW \to 4$ leptons. The heavy Higgs ($h_2$) graphs define the signal process, 
which interferes with the light Higgs ($h_1$) graphs (a,b).  They also interfere
with the gluon-induced continuum background graphs (c,d).}
\end{figure}
In section \ref{sec:results} we present results calculated with a new
version of \textsf{gg2VV} \cite{Kauer:2012hd,Kauer:2013qba,Kauer:2012ma}, which is publicly available \cite{gg2VVurl}. Representative Feynman graphs for the light 
and heavy Higgs and interfering continuum background processes are shown in figure 
\ref{fig:diagrams}.  The heavy Higgs ($h_2$) graphs define the signal process. 
They interfere with the light Higgs ($h_1$) graphs and with the gluon-induced 
continuum background graphs.
The amplitudes are calculated using a modified (for compatibility only) output of \textsf{FeynArts/FormCalc} \cite{Hahn:2000kx,Hahn:1998yk}, using a custom coded \textsf{UFO} \cite{Degrande:2011ua} model file generated by \textsf{FeynRules} \cite{Alloul:2013bka}. The Higgs boson widths are calculated using \textsf{FeynRules} for consistency. The used width values are given in 
table \ref{tab:widths}.
\begin{table}[tb]
\begin{center}
\begin{tabular}{|c|c|c|c|c|c|}
\cline{3-6}
\multicolumn{2}{c|}{} & $h_1$ & \multicolumn{3}{|c|}{$h_2$}  \\ \cline{2-6}
\multicolumn{1}{c|}{} & $M$ [GeV] & 125 & 300 & 600 & 900 \\ \hline
$\theta=\pi/15$ & $\Gamma$ [GeV] & $4.77358\times 10^{-3}$ & 0.5383 & 6.42445 & 21.4215 \\ \hline
$\theta=\pi/8$ &  $\Gamma$ [GeV] & $4.2577\times 10^{-3}$ &1.70204 & 20.7236 	& 69.1805 	\\ \hline
\end{tabular}
\caption{\label{tab:widths}Widths of the physical Higgs bosons $h_1$ and $h_2$ in the 1-Higgs-Singlet Extension of the SM with mixing angles $\theta=\pi/15$ and $\theta=\pi/8$ as well as $\mu_1= \lambda_1=\lambda_2=0$.}
\end{center}\label{default}
\end{table}
The PDF set MSTW2008LO \cite{Martin:2009iq} with default $\alpha_s$ is used and the CKM matrix is approximated by the unit matrix, which causes a negligible error \cite{Kauer:2012ma}. 
As input parameters, we use the specification of the 
LHC Higgs Cross Section Working Group in App.~A of ref.\ \cite{Dittmaier:2011ti}
with $G_\mu$ scheme and LO weak boson widths for consistency.
More specifically, $M_W = 80.398$ GeV, $M_Z = 91.1876$ GeV, $\Gamma_W = 2.141$ GeV, $
\Gamma_Z = 2.4952$ GeV, $M_t = 172.5$ GeV, $M_b = 4.75$ GeV, $G_F = 1.16637\cdot 10^{-5}$ GeV$^{-2}$ are used. Finite top and bottom quark mass effects 
are included.  Lepton masses are neglected.
A fixed-width Breit-Wigner propagator is employed for the weak bosons 
and the Higgs boson. 
The width parameter of the complex pole of the Higgs propagator
is defined in eq.\ (16) of ref.\ \cite{Goria:2011wa}.
The box graphs shown in figure \ref{fig:diagrams}(c,d)
are affected by numerical instabilities when Gram determinants 
approach zero.  In these critical phase space regions the amplitude 
is evaluated in quadruple precision.  Residual instabilities are eliminated 
by requiring that $p_{T,W}$ and $p_{T,Z}$ are larger than $1$ GeV.
This criterion is also applied to the Higgs amplitudes, which are 
not affected by numerical instabilities, to obtain consistent 
cross section-level results.  The numerical effect of this technical cut
has been shown to be small \cite{Kauer:2012ma,Campbell:2013una}.
Furthermore, minimal selection cuts are applied: $M_{\ell\bar{\ell}} > 4$ GeV and 
$M_{\ell'\bar{\ell}'} > 4$ GeV cuts are applied for the $gg\to Z(\gamma^\ast)Z(\gamma^\ast)\to \ell\bar{\ell}\ell'\bar{\ell}'$ process to eliminate the soft photon singularities. The renormalisation and factorisation scales are set to $M_{VV}/2$ and the 
$pp$ collision energy is $\sqrt{s}=8$ TeV.

The phase space integration is carried out using the multi-channel Monte Carlo 
integration technique \cite{Berends:1994pv}, in which every kinematic structure 
has its own mapping from random variables to the phase space configuration 
such that singularities or peaks in the amplitude are compensated, and the 
inverse Jacobi determinants of all mappings are summed to give the inverse weight at
each phase space point. This approach has the advantage of a straightforward
systematic extension from 
the SM to two-Higgs models: an extra channel with a mapping for the heavy Higgs resonance is added. The multi-channel technique has been implemented in the new version of \textsf{gg2VV}, and has been tested thoroughly.  Each mapping was phase space integrated 
individually to check that the result matches the analytically known phase space volume for massless final state particles. 
Cross sections for the continuum background and $h_1$ only contributions\footnote{without mixing, i.e.\ $\theta=0$} to the processes considered here were found to be in 
agreement with the results of ref.\ \cite{Kauer:2013qba}, which were calculated using a previous version of \textsf{gg2VV} with a different phase space implementation based on a decomposition into sections.
Furthermore,
results for similar processes calculated using the same code show excellent 
agreement with a fully independent implementation \cite{Kauer:2015dma}. 


\section{Results\label{sec:results}}

In this section we present integrated and differential cross section-level
results for the $h_2$ signal ($S$) and its interference ($I$) at the LHC for the 
processes
\begin{equation}
g g\ (\to \{h_1,h_2\}) \to Z(\gamma^\ast)Z(\gamma^\ast) \to \ell\bar{\ell}\ell'\bar{\ell}'
\label{procZZ}
\end{equation}
and
\begin{equation}
g g\ (\to \{h_1,h_2\}) \to W^-W^+ \to \ell\bar{\nu}\bar{\ell}'\nu' 
\label{procWW}
\end{equation}
with input parameters, settings and cuts as described in section \ref{sec:calc}.
The following notation is used:
\ea{ S &\sim \left|\calM_{h2}\right|^2\\I_{h1} &\sim 2\,\mathrm{Re}(\calM_{h2}^\ast\,\calM_{h1})\\ I_{bkg} &\sim 2\,\mathrm{Re}(\calM_{h2}^\ast\,\calM_{bkg})\\I_{full} &= I_{h1}+I_{bkg}\\R_i 
&= \frac{S+I_i}{S}\,.}
The interference of the heavy Higgs signal with the light Higgs and continuum background is given separately. We also give the combined interference to illustrate the overall effect.  The ratios $R_{h1}$, $R_{bkg}$ and $R_{full}$ illustrate the relative change of the heavy Higgs signal due to interference with the 
light Higgs and continuum background amplitude contributions.
Integrated results for processes \ref{procZZ} and \ref{procWW} are shown in tables \ref{tab:ZZ}--\ref{tab:WWalt}.
As illustrated by the differential distributions shown below, a $|M_{VV}-M_{h2}| < \Gamma_{h2}$ window cut is an effective means to eliminate or mitigate the 
interference.\footnote{%
For process \protect\ref{procWW}, an invariant $M_{WW}$ cut cannot be applied experimentally. However, a transverse mass cut is feasible.}
Therefore, integrated results with window cut are presented in tables 
\ref{tab:ZZwindow}--\ref{tab:WWwindowalt}.
\begin{table}[tb]
\vspace*{-0cm}
\footnotesize
\renewcommand{\arraystretch}{1.2}
\begin{tabular}{|l|c|ccc|ccc|}
\cline{1-2}
\multicolumn{2}{|c|}{$ gg \to h_2\to ZZ \to \ell\bar{\ell} \ell' \bar{\ell}' $} & \multicolumn{3}{|c}{} \\ 
  \multicolumn{2}{|c|}{$\sigma$ [fb], $pp$, $\sqrt{s}=8$ TeV} & \multicolumn{3}{|c}{} \\ 
\cline{3-8}
  \multicolumn{2}{|c|}{min.\ cuts, $\theta=\pi/15$} &
\multicolumn{3}{c|}{interference} & \multicolumn{3}{c|}{ratio} \\ 
\hline
 $ M_{h2}$  [GeV]  & $S$ & $I_{h1}$  & $I_{bkg}$ & $I_{full}$  &$R_{h1}$ & $R_{bkg}$ & $R_{full}$ \\ \hline
300	&	0.033453(7)	&	0.00392(2)	&	0.00105(2)	&	0.00499(2)	&	1.1171(6)	&	1.0315(7)	&	1.1492(6)	\\ \hline
600	&	0.005223(4)	&	-0.001738(8)	&	0.001730(9)	&	-9(4)e-06	&	0.667(2)	&	1.331(2)	&	0.998(2)	\\ \hline
900	&	0.0005088(4)	&	-0.001151(2)	&	0.001043(3)	&	-0.0001092(9)	&	-1.263(5)	&	3.049(5)	&	0.785(2)	\\ \hline
\end{tabular}\\[0cm] 
\caption{\label{tab:ZZ}
Cross sections for $ g g\ (\to \{h_1,h_2\}) \to ZZ \to \ell \bar{\ell}  \ell' \bar{\ell}'$ in $pp$ collisions at $\sqrt{s}=8$ TeV at loop-induced leading order 
in the 1-Higgs-Singlet Extension of the SM with $M_{h1} = 125$ GeV, $M_{h2} = 300, 600, 900$ GeV and mixing angle $\theta=\pi/15$.
Results for the heavy Higgs ($h_2$) signal ($S$) and its interference with the light Higgs 
($I_{h1}$) and the continuum background ($I_{bkg}$) and the full interference 
($I_{full}$) are given.
The ratio $R_i = (S+I_i)/S$ illustrates the relative change of the heavy Higgs 
signal due to interference with the light Higgs and continuum background amplitude 
contributions.
Minimal cuts are applied (see main text).
Cross sections are given for a single lepton flavour combination.  
The integration error is displayed in brackets.
} 
\end{table}
%
\begin{table}[tb]
\vspace*{-0cm}
\footnotesize
\renewcommand{\arraystretch}{1.2}
\begin{tabular}{|l|c|ccc|ccc|}
\cline{1-2}
\multicolumn{2}{|c|}{$ gg \to h_2\to ZZ \to \ell\bar{\ell} \ell' \bar{\ell}' $} & \multicolumn{3}{|c}{} \\ 
  \multicolumn{2}{|c|}{$\sigma$ [fb], $pp$, $\sqrt{s}=8$ TeV} & \multicolumn{3}{|c}{} \\ 
\cline{3-8}
  \multicolumn{2}{|c|}{min.\ cuts, $\theta=\pi/8$} &
\multicolumn{3}{c|}{interference} & \multicolumn{3}{c|}{ratio} \\ 
\hline
 $ M_{h2}$  [GeV]  & $S$ & $I_{h1}$  & $I_{bkg}$ & $I_{full}$  &$R_{h1}$ & $R_{bkg}$ & $R_{full}$ \\ \hline
300	&	0.12209(9)	&	0.0119(1)	&	0.00358(5)	&	0.01545(4)	&	1.097(2)	&	1.029(2)	&	1.127(2)	\\ \hline
600	&	0.01821(2)	&	-0.00498(2)	&	0.00568(2)	&	0.000694(8)	&	0.727(2)	&	1.312(2)	&	1.038(2)	\\ \hline
900	&	0.001781(2)	&	-0.003277(5)	&	0.003396(5)	&	0.000118(3)	&	-0.840(3)	&	2.906(4)	&	1.066(2)	\\ \hline
\end{tabular}\\[0cm] 
\caption{\label{tab:ZZalt}
Cross sections for $ g g\ (\to \{h_1,h_2\}) \to ZZ \to \ell \bar{\ell}  \ell' \bar{\ell}'$ in $pp$ collisions in the 1-Higgs-Singlet Extension of the SM with mixing angle $\theta=\pi/8$.  Other details as in table \protect\ref{tab:ZZ}.
} 
\end{table}
%
\begin{table}[tb]
\vspace*{-0cm}
\footnotesize
\renewcommand{\arraystretch}{1.2}
\begin{tabular}{|l|c|ccc|ccc|}
\cline{1-2}
\multicolumn{2}{|c|}{$gg \to  h_2\to W^-W^+\to \ell\bar{\nu} \bar{\ell}'\nu' $} & \multicolumn{3}{|c}{} \\ 
  \multicolumn{2}{|c|}{$\sigma$ [fb], $pp$, $\sqrt{s}=8$ TeV} & \multicolumn{3}{|c}{} \\ 
\cline{3-8}
  \multicolumn{2}{|c|}{min.\ cuts, $\theta=\pi/15$} &
\multicolumn{3}{c|}{interference} & \multicolumn{3}{c|}{ratio} \\ 
\hline
 $ M_{h2}$  [GeV]  & $S$ & $I_{h1}$  & $I_{bkg}$ & $I_{full}$  &$R_{h1}$ & $R_{bkg}$ & $R_{full}$ \\ \hline
300	&	0.3752(3)	&	0.0391(9)	&	-0.0132(7)	&	0.0254(5)	&	1.104(3)	&	0.965(3)	&	1.068(2)	\\ \hline
600	&	0.05380(4)	&	-0.0191(2)	&	0.0289(2)	&	0.00957(8)	&	0.645(3)	&	1.536(4)	&	1.178(2)	\\ \hline
900	&	0.005149(4)	&	-0.01217(6)	&	0.01519(4)	&	0.00300(3)	&	-1.36(2)	&	3.950(9)	&	1.582(5)	\\ \hline
\end{tabular}\\[0cm] 
\caption{\label{tab:WW}
Cross sections for $ g g\ (\to \{h_1,h_2\}) \to W^-W^+ \to \ell \bar{\nu} \bar{\ell}' \nu'$ in $pp$ collisions at $\sqrt{s}=8$ TeV  
in the 1-Higgs-Singlet Extension of the SM with $M_{h1} = 125$ GeV, $M_{h2} = 300, 600, 900$ GeV and mixing angle $\theta=\pi/15$.
Other details as in table \protect\ref{tab:ZZ}.} 
\end{table}
%
\begin{table}[tb]
\vspace*{-0cm}
\footnotesize
\renewcommand{\arraystretch}{1.2}
\begin{tabular}{|l|c|ccc|ccc|}
\cline{1-2}
\multicolumn{2}{|c|}{$gg \to  h_2\to W^-W^+\to \ell\bar{\nu} \bar{\ell}'\nu' $} & \multicolumn{3}{|c}{} \\ 
  \multicolumn{2}{|c|}{$\sigma$ [fb], $pp$, $\sqrt{s}=8$ TeV} & \multicolumn{3}{|c}{} \\ 
\cline{3-8}
  \multicolumn{2}{|c|}{min.\ cuts, $\theta=\pi/8$} &
\multicolumn{3}{c|}{interference} & \multicolumn{3}{c|}{ratio} \\ 
\hline
 $ M_{h2}$  [GeV]  & $S$ & $I_{h1}$  & $I_{bkg}$ & $I_{full}$  &$R_{h1}$ & $R_{bkg}$ & $R_{full}$ \\ \hline
300	&	1.368(2)	&	0.118(2)	&	-0.045(2)	&	0.0712(9)	&	1.086(2)	&	0.967(2)	&	1.052(2)	\\ \hline
															
600	&	0.1875(2)	&	-0.0548(3)	&	0.0940(4)	&	0.0389(3)	&	0.708(2)	&	1.501(3)	&	1.207(2)	\\ \hline
															
900	&	0.01806(2)	&	-0.03467(8)	&	0.0495(2)	&	0.01478(7)	&	-0.920(5)	&	3.742(7)	&	1.818(5)	\\ \hline
\end{tabular}\\[0cm] 
\caption{\label{tab:WWalt}
Cross sections for $ g g\ (\to \{h_1,h_2\}) \to W^-W^+ \to \ell \bar{\nu} \bar{\ell}' \nu'$ in $pp$ collisions in the 1-Higgs-Singlet Extension of the SM with mixing angle $\theta=\pi/8$.
Other details as in table \protect\ref{tab:WW}.} 
\end{table}
%
\begin{table}[tb]
 \footnotesize
 \renewcommand{\arraystretch}{1.2}
 \begin{tabular}{|l|c|ccc|ccc|}
\cline{1-2}
\multicolumn{2}{|c|}{$ gg\to h_2\to ZZ \to \ell\bar{\ell} \ell' \bar{\ell}' $} & \multicolumn{3}{|c}{} \\ 
  \multicolumn{2}{|c|}{$\sigma$ [fb], $pp$, $\sqrt{s}=8$ TeV} & \multicolumn{3}{|c}{} \\ 
  \multicolumn{2}{|c|}{min.\ cuts \& $|M_{VV}-M_{h2}| < \Gamma_{h2}$} & \multicolumn{3}{|c}{} \\ 
\cline{3-8}
  \multicolumn{2}{|c|}{$\theta=\pi/15$} &
\multicolumn{3}{c|}{interference} & \multicolumn{3}{c|}{ratio} \\ 
\hline
 $ M_{h2}$  [GeV]  & $S$ & $I_{h1}$  & $I_{bkg}$ & $I_{full}$  &$R_{h1}$ & $R_{bkg}$ & $R_{full}$ \\ \hline
300	&	0.02352(2)	&	3.8(4)e-06	&	0.001583(3)	&	0.001586(3)	&	1.000(2)	&	1.067(2)	&	1.067(2)	\\ \hline
600	&	0.003719(4)	&	-1.7(2)e-05	&	0.000288(2)	&	0.000271(2)	&	0.995(2)	&	1.077(2)	&	1.073(2)	\\ \hline
900	&	0.0003606(3)	&	-1.35(2)e-05	&	8.56(3)e-05	&	7.21(4)e-05	&	0.963(2)	&	1.237(2)	&	1.200(2)	\\ \hline
\end{tabular}\\[0cm]
\caption{\label{tab:ZZwindow}
Cross sections for $ g g\ (\to \{h_1,h_2\}) \to ZZ \to \ell \bar{\ell}  \ell' \bar{\ell}'$ in $pp$ collisions at $\sqrt{s}=8$ TeV  
in the 1-Higgs-Singlet Extension of the SM with $M_{h1} = 125$ GeV, $M_{h2} = 300, 600, 900$ GeV and mixing angle $\theta=\pi/15$.
An additional window cut  $\left|M_{ZZ} - M_{h2}\right| < \Gamma_{h2}$
is applied. Other details as in table \protect\ref{tab:ZZ}.}
\end{table}
%
\begin{table}[tb]
 \footnotesize
 \renewcommand{\arraystretch}{1.2}
 \begin{tabular}{|l|c|ccc|ccc|}
\cline{1-2}
\multicolumn{2}{|c|}{$ gg\to h_2\to ZZ \to \ell\bar{\ell} \ell' \bar{\ell}' $} & \multicolumn{3}{|c}{} \\ 
  \multicolumn{2}{|c|}{$\sigma$ [fb], $pp$, $\sqrt{s}=8$ TeV} & \multicolumn{3}{|c}{} \\ 
  \multicolumn{2}{|c|}{min.\ cuts \& $|M_{VV}-M_{h2}| < \Gamma_{h2}$} & \multicolumn{3}{|c}{} \\ 
\cline{3-8}
  \multicolumn{2}{|c|}{$\theta=\pi/8$} &
\multicolumn{3}{c|}{interference} & \multicolumn{3}{c|}{ratio} \\ 
\hline
 $ M_{h2}$  [GeV]  & $S$ & $I_{h1}$  & $I_{bkg}$ & $I_{full}$  &$R_{h1}$ & $R_{bkg}$ & $R_{full}$ \\ \hline
300	&	0.08537(8)	&	3.6(4)e-05	&	0.005371(9)	&	0.00541(1)	&	1.000(2)	&	1.063(2)	&	1.063(2)	\\ \hline
600	&	0.01323(2)	&	-0.000174(4)	&	0.001058(4)	&	0.000884(6)	&	0.987(2)	&	1.080(2)	&	1.067(2)	\\ \hline
900	&	0.001283(1)	&	-0.0001316(9)	&	0.000373(1)	&	0.000241(2)	&	0.897(2)	&	1.290(2)	&	1.188(2)	\\ \hline
\end{tabular}\\[0cm]
\caption{\label{tab:ZZwindowalt}
Cross sections for $ g g\ (\to \{h_1,h_2\}) \to ZZ \to \ell \bar{\ell}  \ell' \bar{\ell}'$ in $pp$ collisions in the 1-Higgs-Singlet Extension of the SM with mixing angle $\theta=\pi/8$.  Other details as in table \protect\ref{tab:ZZwindow}.}
\end{table}
%
\begin{table}[tb]
 \footnotesize
 \renewcommand{\arraystretch}{1.2}
 \begin{tabular}{|l|c|ccc|ccc|}
\cline{1-2}
\multicolumn{2}{|c|}{$ gg \to h_2\to W^-W^+\to \ell\bar{\nu} \bar{\ell}'\nu' $} & \multicolumn{3}{|c}{} \\ 
  \multicolumn{2}{|c|}{$\sigma$ [fb], $pp$, $\sqrt{s}=8$ TeV} & \multicolumn{3}{|c}{} \\ 
  \multicolumn{2}{|c|}{min.\ cuts \& $|M_{VV}-M_{h2}| < \Gamma_{h2}$} & \multicolumn{3}{|c}{} \\ 
\cline{3-8}
  \multicolumn{2}{|c|}{$\theta=\pi/15$} &
\multicolumn{3}{c|}{interference} & \multicolumn{3}{c|}{ratio} \\ 
\hline
 $ M_{h2}$  [GeV]  & $S$ & $I_{h1}$  & $I_{bkg}$ & $I_{full}$  &$R_{h1}$ & $R_{bkg}$ & $R_{full}$ \\ \hline
300	&	0.3352(3)	&	3.8(6)e-05	&	0.00959(6)	&	0.00963(7)	&	1.000(2)	&	1.029(2)	&	1.029(2)	\\ \hline
600	&	0.04859(5)	&	-0.000188(4)	&	0.00419(3)	&	0.00401(3)	&	0.996(2)	&	1.086(2)	&	1.082(2)	\\ \hline
900	&	0.004635(5)	&	-0.000137(3)	&	0.000929(5)	&	0.000792(5)	&	0.970(2)	&	1.200(2)	&	1.171(2)	\\ \hline
\end{tabular}\\[0cm]
\caption{\label{tab:WWwindow}
Cross sections for $ g g\ (\to \{h_1,h_2\}) \to W^-W^+ \to \ell \bar{\nu} \bar{\ell}' \nu'$ in $pp$ collisions at $\sqrt{s}=8$ TeV  
in the 1-Higgs-Singlet Extension of the SM with $M_{h1} = 125$ GeV, $M_{h2} = 300, 600, 900$ GeV and mixing angle $\theta=\pi/15$.
An additional window cut  $\left|M_{WW} - M_{h2}\right| < \Gamma_{h2}$
is applied.
Other details as in table \protect\ref{tab:ZZ}.}
\end{table}
%
\begin{table}[tb]
 \footnotesize
 \renewcommand{\arraystretch}{1.2}
 \begin{tabular}{|l|c|ccc|ccc|}
\cline{1-2}
\multicolumn{2}{|c|}{$ gg \to h_2\to W^-W^+\to \ell\bar{\nu} \bar{\ell}'\nu' $} & \multicolumn{3}{|c}{} \\ 
  \multicolumn{2}{|c|}{$\sigma$ [fb], $pp$, $\sqrt{s}=8$ TeV} & \multicolumn{3}{|c}{} \\ 
  \multicolumn{2}{|c|}{min.\ cuts \& $|M_{VV}-M_{h2}| < \Gamma_{h2}$} & \multicolumn{3}{|c}{} \\ 
\cline{3-8}
  \multicolumn{2}{|c|}{$\theta=\pi/8$} &
\multicolumn{3}{c|}{interference} & \multicolumn{3}{c|}{ratio} \\ 
\hline
 $ M_{h2}$  [GeV]  & $S$ & $I_{h1}$  & $I_{bkg}$ & $I_{full}$  &$R_{h1}$ & $R_{bkg}$ & $R_{full}$ \\ \hline
300	&	0.9578(9)	&	0.00034(2)	&	0.0324(2)	&	0.0329(2)	&	1.000(2)	&	1.034(2)	&	1.034(2)	\\ \hline
600	&	0.1361(2)	&	-0.00184(2)	&	0.01578(6)	&	0.01394(3)	&	0.987(2)	&	1.116(2)	&	1.102(2)	\\ \hline
900	&	0.01298(1)	&	-0.001340(7)	&	0.00429(2)	&	0.002952(7)	&	0.897(2)	&	1.331(2)	&	1.227(2)	\\ \hline
\end{tabular}\\[0cm]
\caption{\label{tab:WWwindowalt}
Cross sections for $ g g\ (\to \{h_1,h_2\}) \to W^-W^+ \to \ell \bar{\nu} \bar{\ell}' \nu'$ in $pp$ collisions in the 1-Higgs-Singlet Extension of the SM with mixing angle $\theta=\pi/8$.
Other details as in table \protect\ref{tab:WWwindow}.}
\end{table}

\begin{figure}[tb]
\centering
\includegraphics[width = 0.6\textwidth, clip=true]{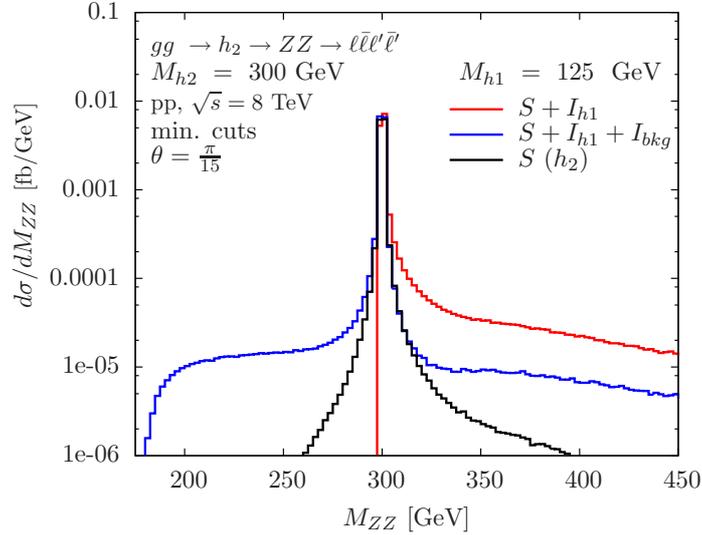}
\caption{\label{fig:ZZ300} 
Invariant $ZZ$ mass distributions for $ g g\ (\to \{h_1,h_2\}) \to ZZ \to \ell \bar{\ell}  \ell' \bar{\ell}'$ in $pp$ collisions at $\sqrt{s}=8$ TeV at loop-induced leading order in the 1-Higgs-Singlet Extension of the SM with $M_{h1} = 125$ GeV, $M_{h2} = 300$ GeV and mixing angle $\theta=\pi/15$.
Results for the heavy Higgs ($h_2$) signal ($S$) and including interference with the light Higgs 
($S+I_{h1}$) and the continuum background ($S+I_{h1}+I_{bkg}$) are shown.
Minimal cuts are applied (see main text).
Other details as in table \protect\ref{tab:ZZ}.
}
\end{figure}

\begin{figure}[tb]
\centering
\includegraphics[width = 0.6\textwidth, clip=true]{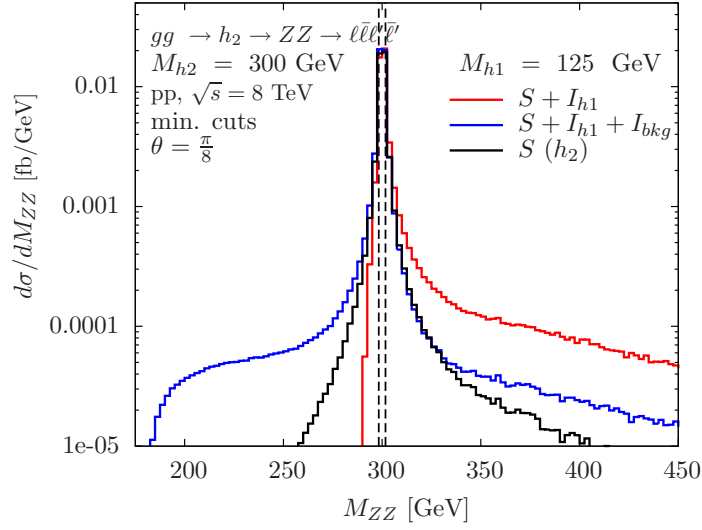}
\caption{\label{fig:ZZ300alt} 
Invariant $ZZ$ mass distributions for $ g g\ (\to \{h_1,h_2\}) \to ZZ \to \ell \bar{\ell}  \ell' \bar{\ell}'$ in $pp$ collisions in the 1-Higgs-Singlet Extension of the SM with mixing angle $\theta=\pi/8$.
Vertical dashed lines are shown at $M_{VV}=M_{h2} \pm\Gamma_{h2}$.
Other details as in figure \protect\ref{fig:ZZ300}.
}
\end{figure}

\begin{figure}[tb]
\centering
\includegraphics[width=0.6\textwidth, clip=true]{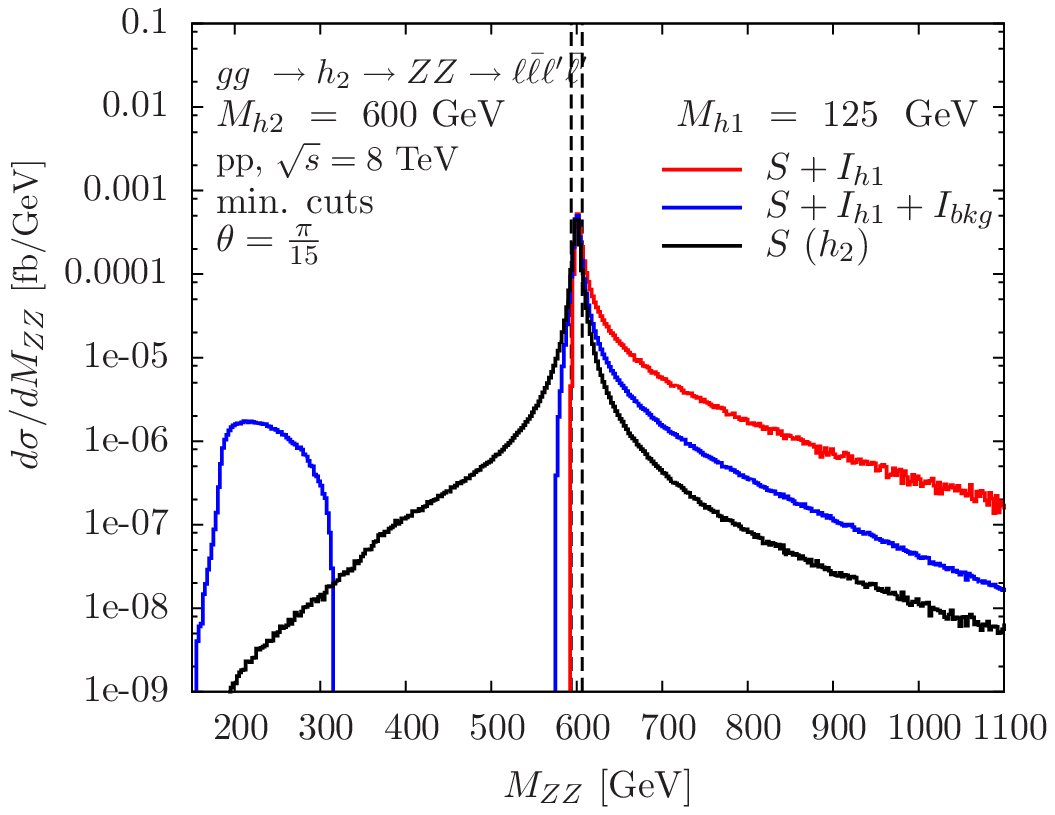}
\caption{\label{fig:ZZ600}
Invariant $ZZ$ mass distributions for $ g g\ (\to \{h_1,h_2\}) \to ZZ \to \ell \bar{\ell}  \ell' \bar{\ell}'$ in $pp$ collisions at $\sqrt{s}=8$ TeV in the 1-Higgs-Singlet Extension of the SM with $M_{h1} = 125$ GeV, $M_{h2} = 600$ GeV and mixing angle $\theta=\pi/15$.
Other details as in figure \ref{fig:ZZ300}.}
\end{figure}

\begin{figure}[tb]
\centering
\includegraphics[width=0.6\textwidth, clip=true]{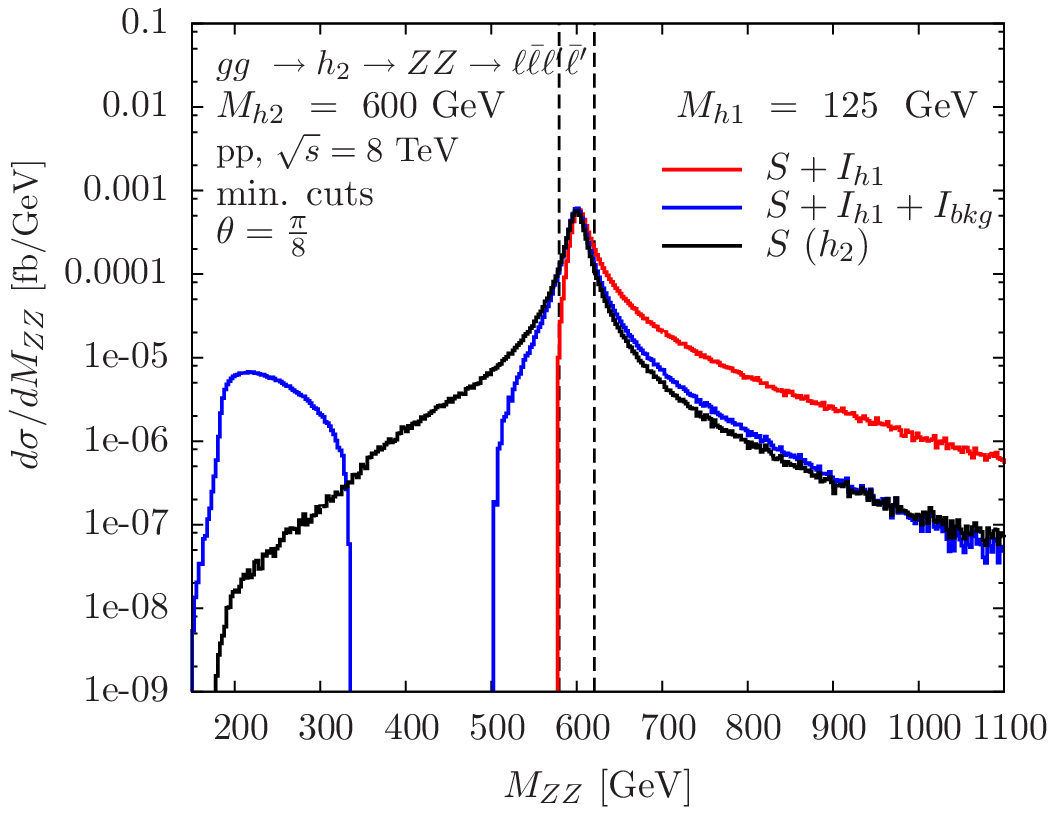}
\caption{\label{fig:ZZ600alt}
Invariant $ZZ$ mass distributions for $ g g\ (\to \{h_1,h_2\}) \to ZZ \to \ell \bar{\ell}  \ell' \bar{\ell}'$ in $pp$ collisions in the 1-Higgs-Singlet Extension of the SM with mixing angle $\theta=\pi/8$.
Other details as in figure \ref{fig:ZZ600}.}
\end{figure}

\begin{figure}[tb]
\centering
\includegraphics[width=0.6\textwidth, clip=true]{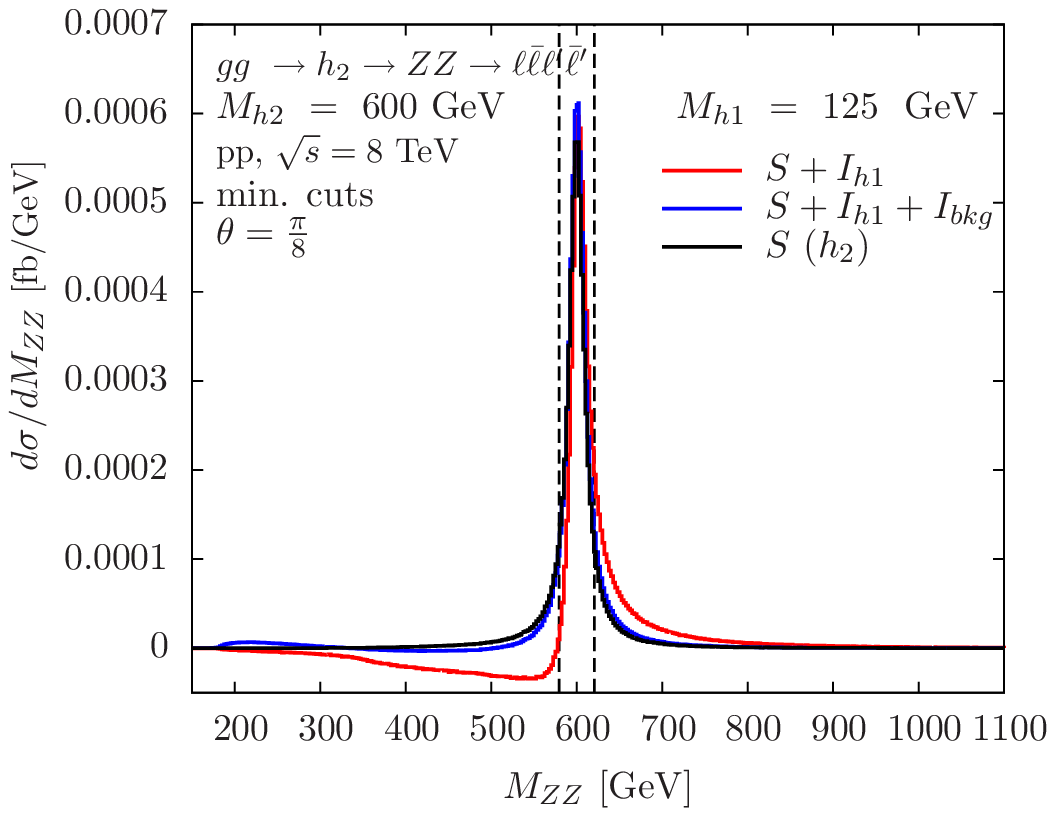}
\caption{\label{fig:ZZ600altlinear} 
Invariant $ZZ$ mass distributions for $ g g\ (\to \{h_1,h_2\}) \to ZZ \to \ell \bar{\ell}  \ell' \bar{\ell}'$ in $pp$ collisions in the 1-Higgs-Singlet Extension of the SM with mixing angle $\theta=\pi/8$.  As figure \ref{fig:ZZ600alt}, but with linear $d\sigma/dM_{ZZ}$ scale, 
to illustrate negative $S+I_{h1}$ and $S+I_{h1}+I_{bkg}$.}
\end{figure}

\begin{figure}[tb]
\centering
\includegraphics[width=0.6\textwidth, clip=true]{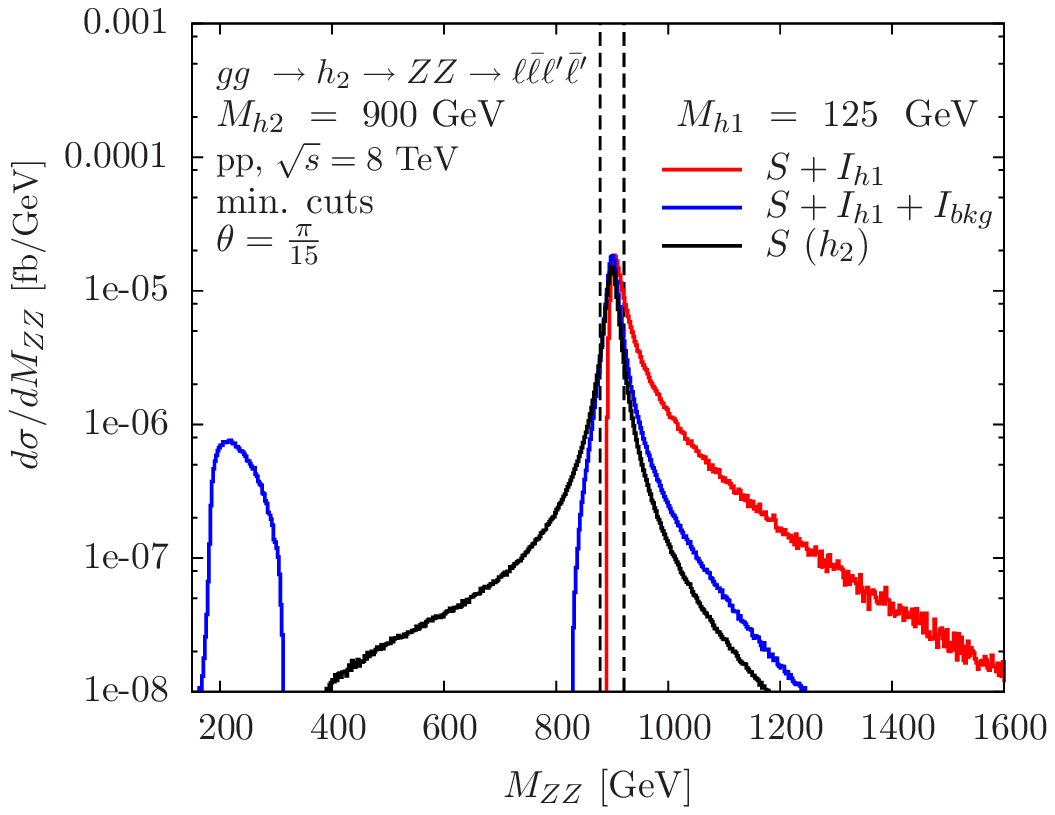}
\caption{\label{fig:ZZ900} 
Invariant $ZZ$ mass distributions for $ g g\ (\to \{h_1,h_2\}) \to ZZ \to \ell \bar{\ell}  \ell' \bar{\ell}'$ in $pp$ collisions at $\sqrt{s}=8$ TeV in the 1-Higgs-Singlet Extension of the SM with $M_{h1} = 125$ GeV, $M_{h2} = 900$ GeV and mixing angle $\theta=\pi/15$.
Other details as in figure \ref{fig:ZZ300}.}
\end{figure}

\begin{figure}[tb]
\centering
\includegraphics[width=0.6\textwidth, clip=true]{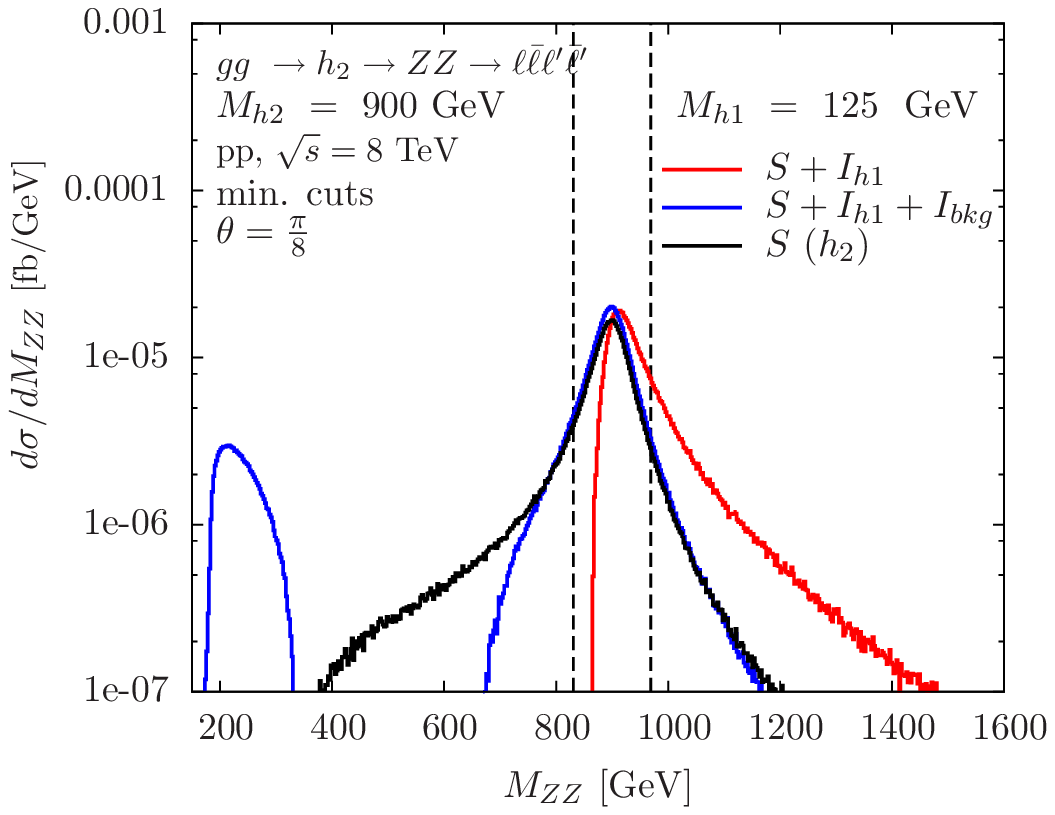}
\caption{\label{fig:ZZ900alt} 
Invariant $ZZ$ mass distributions for $ g g\ (\to \{h_1,h_2\}) \to ZZ \to \ell \bar{\ell}  \ell' \bar{\ell}'$ in $pp$ collisions in the 1-Higgs-Singlet Extension of the SM with mixing angle $\theta=\pi/8$.
Other details as in figure \ref{fig:ZZ900}.}
\end{figure}

\begin{figure}[tb]
\centering
\includegraphics[width=0.6\textwidth, clip=true]{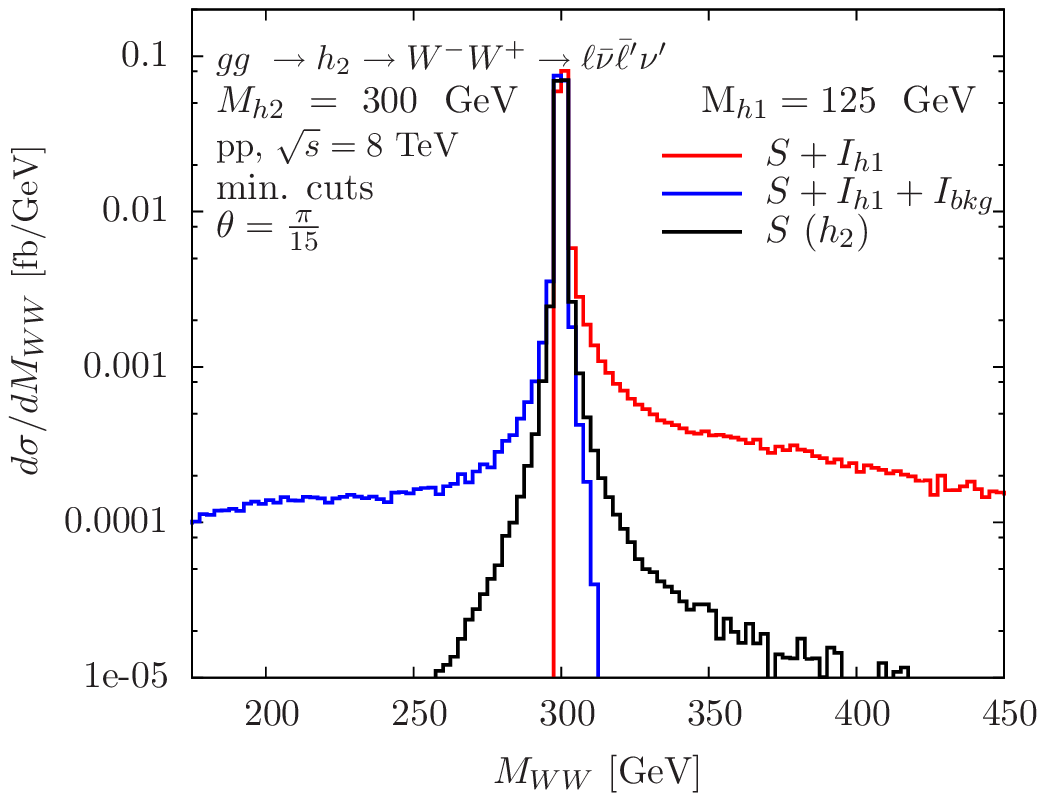}
\caption{\label{fig:WW300} 
Invariant $WW$ mass distributions for $ g g\ (\to \{h_1,h_2\}) \to W^-W^+ \to \ell \bar{\nu} \bar{\ell}' \nu'$ in $pp$ collisions at $\sqrt{s}=8$ TeV in the 1-Higgs-Singlet Extension of the SM with $M_{h1} = 125$ GeV, $M_{h2} = 300$ GeV and mixing angle $\theta=\pi/15$.
Other details as in figure \ref{fig:ZZ300}.
}
\end{figure}

\begin{figure}[tb]
\centering
\includegraphics[width=0.6\textwidth, clip=true]{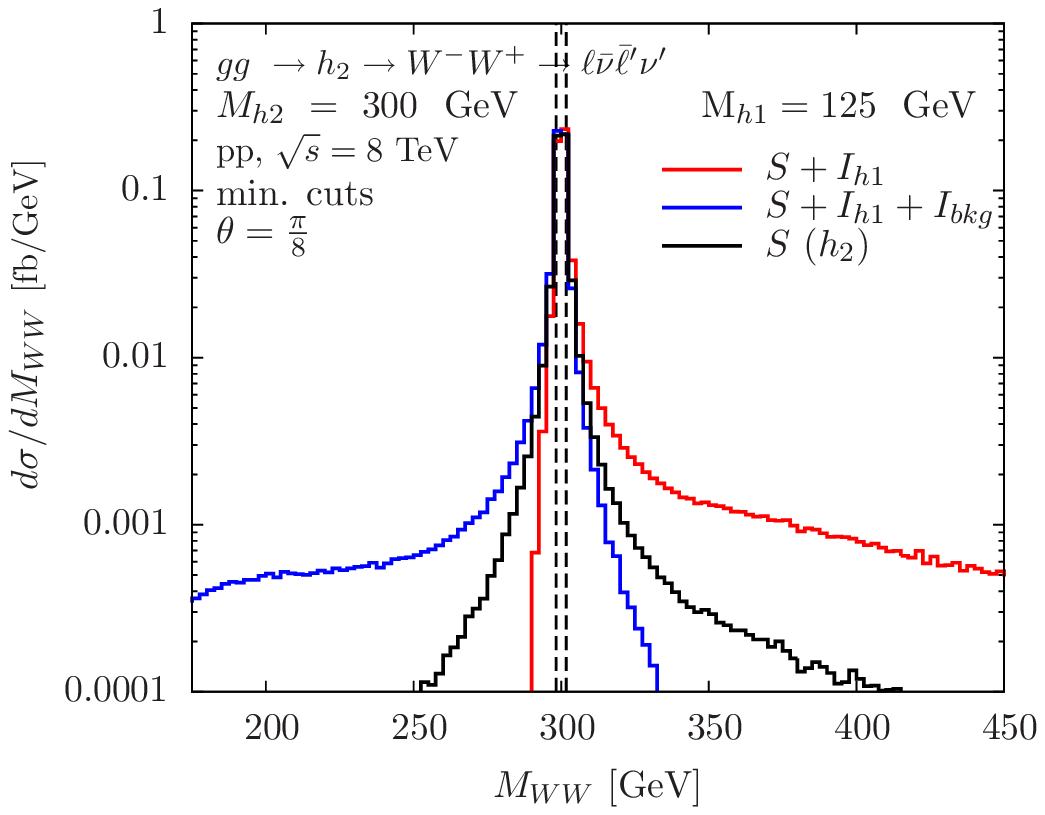}
\caption{\label{fig:WW300alt} 
Invariant $WW$ mass distributions for $ g g\ (\to \{h_1,h_2\}) \to W^-W^+ \to \ell \bar{\nu} \bar{\ell}' \nu'$ in $pp$ collisions in the 1-Higgs-Singlet Extension of the SM with mixing angle $\theta=\pi/8$.
Other details as in figure \ref{fig:WW300}.
}
\end{figure}

\begin{figure}[tb]
\centering
\includegraphics[width=0.6\textwidth, clip=true]{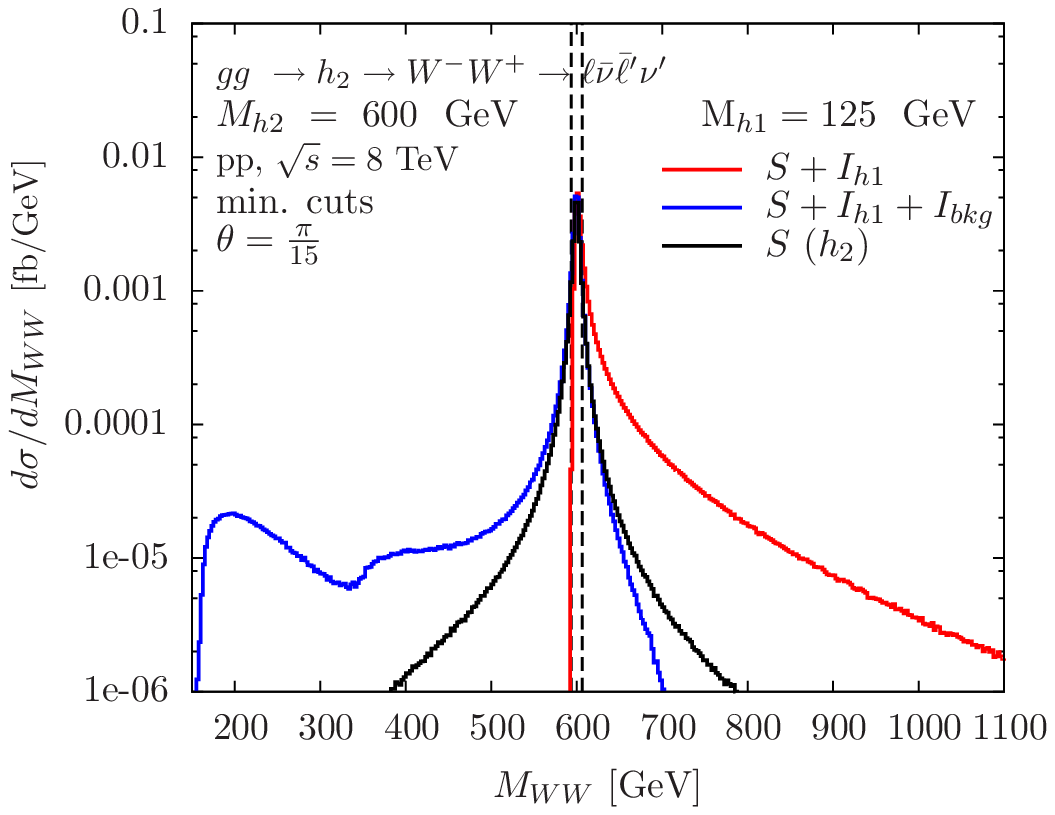}
\caption{\label{fig:WW600}
Invariant $WW$ mass distributions for $ g g\ (\to \{h_1,h_2\}) \to W^-W^+ \to \ell \bar{\nu} \bar{\ell}' \nu'$ in $pp$ collisions at $\sqrt{s}=8$ TeV in the 1-Higgs-Singlet Extension of the SM with $M_{h1} = 125$ GeV, $M_{h2} = 600$ GeV and mixing angle $\theta=\pi/15$.
Other details as in figure \ref{fig:ZZ300}.
}
\end{figure}

\begin{figure}[tb]
\centering
\includegraphics[width=0.6\textwidth, clip=true]{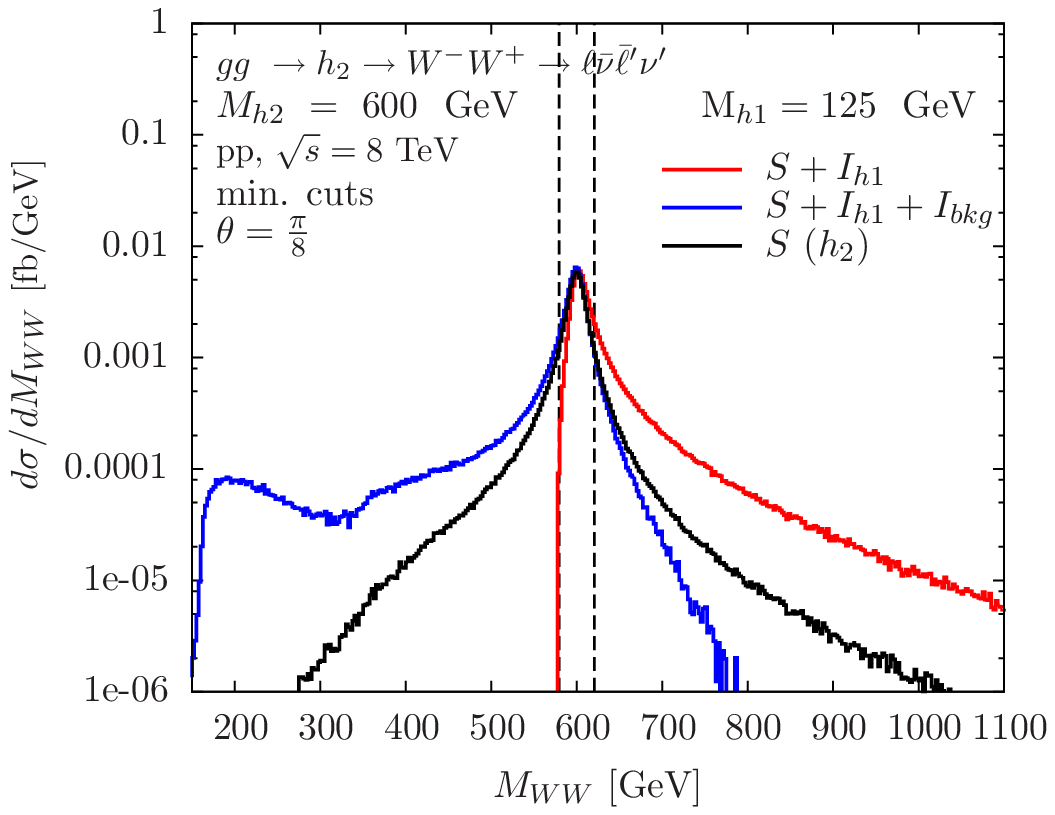}
\caption{\label{fig:WW600alt}
Invariant $WW$ mass distributions for $ g g\ (\to \{h_1,h_2\}) \to W^-W^+ \to \ell \bar{\nu} \bar{\ell}' \nu'$ in $pp$ collisions in the 1-Higgs-Singlet Extension of the SM with mixing angle $\theta=\pi/8$.
Other details as in figure \ref{fig:WW600}.
}
\end{figure}

\begin{figure}[tb]
\centering
\includegraphics[width=0.6\textwidth, clip=true]{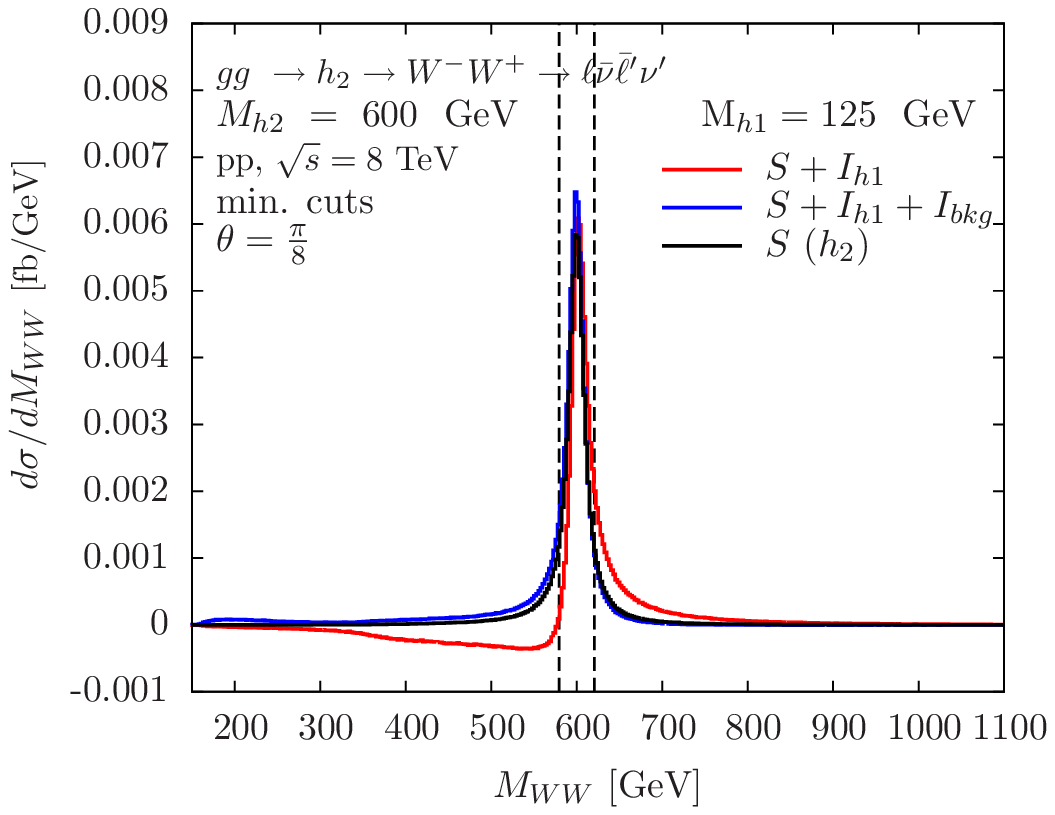}
\caption{\label{fig:WW600altlinear}
Invariant $WW$ mass distributions for $ g g\ (\to \{h_1,h_2\}) \to W^-W^+ \to \ell \bar{\nu} \bar{\ell}' \nu'$ in $pp$ collisions in the 1-Higgs-Singlet Extension of the SM with mixing angle $\theta=\pi/8$.
As figure \ref{fig:WW600alt}, but with linear $d\sigma/dM_{WW}$ scale, 
to illustrate negative $S+I_{h1}$.
}
\end{figure}

\begin{figure}[tb]
\centering
\includegraphics[width=0.6\textwidth, clip=true]{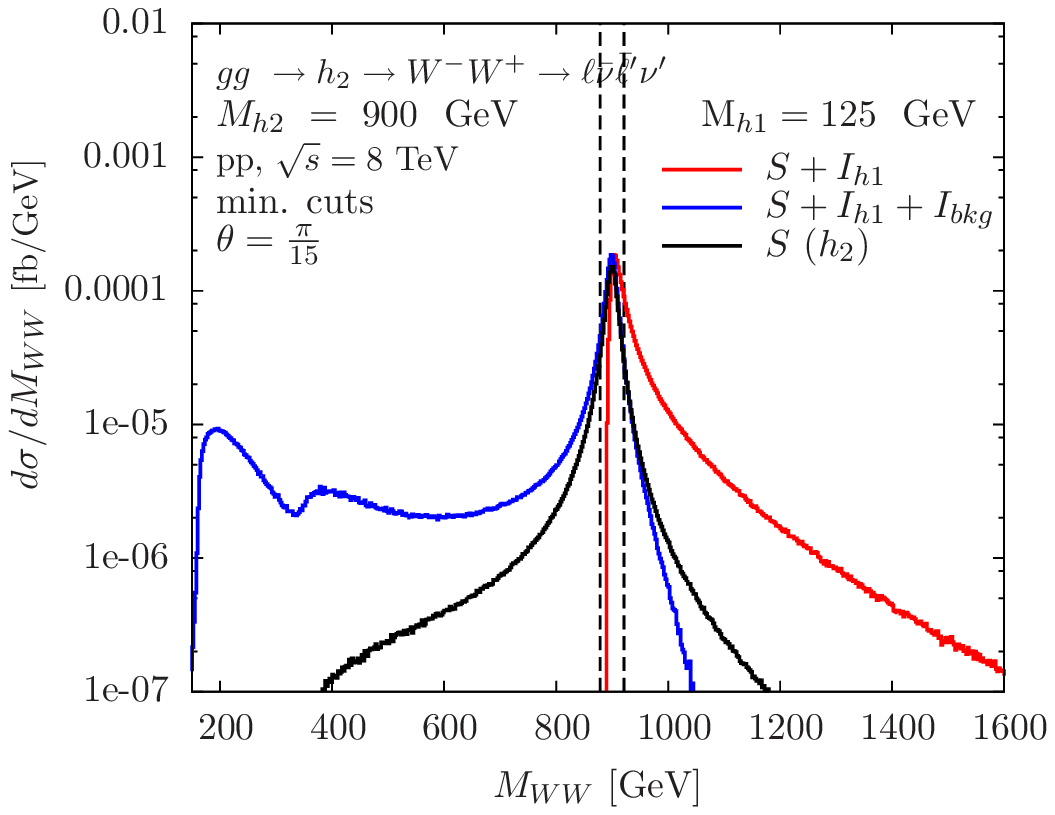}
\caption{\label{fig:WW900}
Invariant $WW$ mass distributions for $ g g\ (\to \{h_1,h_2\}) \to W^-W^+ \to \ell \bar{\nu} \bar{\ell}' \nu'$ in $pp$ collisions at $\sqrt{s}=8$ TeV in the 1-Higgs-Singlet Extension of the SM with $M_{h1} = 125$ GeV, $M_{h2} = 900$ GeV and mixing angle $\theta=\pi/15$.
Other details as in figure \ref{fig:ZZ300}.
}
\end{figure}

\begin{figure}[tb]
\centering
\includegraphics[width=0.6\textwidth, clip=true]{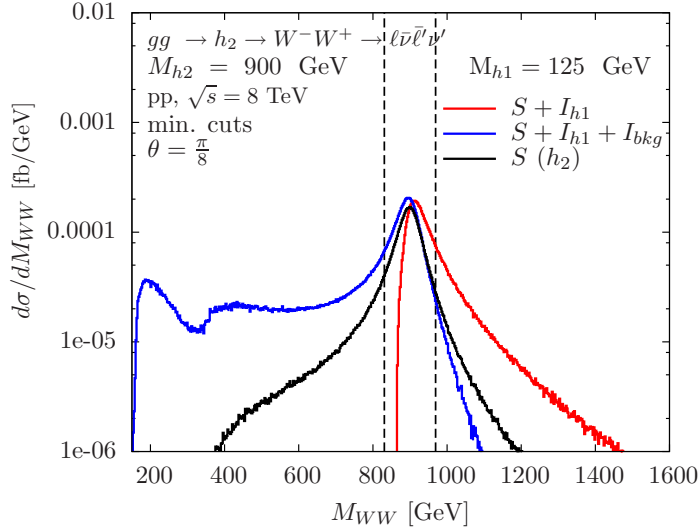}
\caption{\label{fig:WW900alt}
Invariant $WW$ mass distributions for $ g g\ (\to \{h_1,h_2\}) \to W^-W^+ \to \ell \bar{\nu} \bar{\ell}' \nu'$ in $pp$ collisions in the 1-Higgs-Singlet Extension of the SM with mixing angle $\theta=\pi/8$.
Other details as in figure \ref{fig:WW900}.
}
\end{figure}

Corresponding $M_{VV}$ distributions for processes \ref{procZZ} and \ref{procWW}
and $M_{h2} = 300, 600, 900$ GeV
are shown in figures \ref{fig:ZZ300}--\ref{fig:WW900alt}.
Results for the heavy Higgs signal and including interference with the light Higgs 
and the continuum background are displayed.
Where appropriate, vertical dashed lines at $M_{VV}=M_{h2} \pm\Gamma_{h2}$ are used to visualize 
the effect of a $\left|M_{VV} - M_{h2}\right| < \Gamma_{h2}$ window cut.
For invariant $VV$ masses with negative signal plus interference, 
the distributions are shown in figures 
\ref{fig:ZZ600altlinear} and \ref{fig:WW600altlinear}.

As seen in the tables and figures, interference effects increase significantly 
with increasing heavy Higgs mass.  They can range from $\calO(10\%)$ to 
$\calO(1)$ effects for integrated cross sections.  With window cut we 
find that interference effects are mitigated to $\calO(10\%)$ or less.
We note that the heavy Higgs-continuum background interference is 
negative above $M_{h2}$ and positive below $M_{h2}$, while the 
heavy Higgs-light Higgs interference has the opposite behaviour.
Consequently, in the 
heavy Higgs resonance region 
a strong cancellation occurs when both interference contributions are added.
It is therefore essential to take both contributions into account in 
phenomenological and experimental studies.
Despite the occurring cancellation, the full interference is clearly 
non-negligible and modifies
the heavy Higgs line shape.  We find overall $\calO(10\%)$ effects
for integrated cross sections, even if a window cut is applied.
The results for $\theta=\pi/15$ and $\theta=\pi/8$ are in qualitative
agreement.  Relative interference effects show a mild quantitative 
dependence on the mixing angle.


We note that our results for heavy Higgs-light Higgs interference are qualitatively in agreement with those given in ref.\ \cite{Maina:2015ela}, where this interference is considered for $ g g\to \{h_1,h_2\} \to ZZ \to 4\ell$, but in the 1HSM model with an extra $Z_2$ symmetry.

\clearpage

\section{Conclusions\label{sec:sum}}

In the 1-Higgs-Singlet Extension of the SM, the modification 
of the heavy Higgs ($h_2$) signal due to interference with the continuum background 
and the off-shell light Higgs ($h_1$) contribution has been studied for the
$ g g\ (\to \{h_1,h_2\}) \to Z(\gamma^\ast)Z(\gamma^\ast) \to \ell \bar{\ell}  \ell' \bar{\ell}'$
and 
$ g g\ (\to \{h_1,h_2\}) \to W^-W^+ \to \ell \bar{\nu} \bar{\ell}' \nu'$
processes at the LHC.
Interference effects increase significantly 
with increasing heavy Higgs mass.  They can range from $\calO(10\%)$ to 
$\calO(1)$ effects for integrated cross sections.  With a 
$\left|M_{VV} - M_{h2}\right| < \Gamma_{h2}$ window cut, we 
find that interference effects are mitigated to $\calO(10\%)$ or less.
We find that the heavy Higgs-continuum background interference is 
negative above $M_{h2}$ and positive below $M_{h2}$, while the 
heavy Higgs-light Higgs interference has the opposite behaviour.
Consequently, in the 
heavy Higgs resonance region 
a strong cancellation occurs when both interference contributions are added.
It is therefore essential to take both contributions into account in 
phenomenological and experimental studies.
Despite the occurring cancellation, the full interference is clearly 
non-negligible and modifies
the heavy Higgs line shape.  We find overall $\calO(10\%)$ effects
for integrated cross sections, even if a 
window cut is applied
to mitigate the interference effects.
Our calculations have been carried out with a parton-level integrator and event generator, which we have made publicly available.

\acknowledgments
The authors would like to thank A.\ Hadef for a comparison of preliminary 
results during the initial stages of the project and 
S.\ Liebler for providing a preprint.
N.K.\ would like to thank the Galileo Galilei Institute for Theoretical Physics
for hospitality and the INFN for partial support during the preparation of
this paper. C.O.\ would like to thank the Department of Physics, Royal Holloway, 
University of London for supplementary financial support. 
This work was supported by STFC grants ST/J000485/1, ST/J005010/1 
and ST/L000512/1.

\end{document}